# Cellular-based Statistical Model for Mobile Dispersion

Mouhamed Abdulla and Yousef R. Shayan
Department of Electrical and Computer Engineering
Concordia University
Montréal, Québec, Canada
Email: {m_abdull, yshayan}@ece.concordia.ca

*Abstract*—While analyzing mobile systems we often approximate the actual coverage surface and assume an ideal cell shape. In a multi-cellular network, because of its tessellating nature, a hexagon is more preferred than a circular geometry. Despite this reality, perhaps due to the inherent simplicity, only a model for circular based random spreading is available. However, if used, this results an unfair terminal distribution for non-circular contours. Therefore, in this paper we specifically derived an unbiased node density model for a hexagon. We then extended the principle and established stochastic ways to handle sectored cells. Next, based on these mathematical findings, we created a generic modeling tool that can support a complex network with varying position, capacity, size, user density, and sectoring capability. Last, simulation was used to verify the theoretical analysis.

## I. INTRODUCTION

Simulation is almost always used in applied sciences as a smart cost-effective way to learn about a system being designed. In wireless communications, the position and physical separation between interacting nodes is very critical to system parameters such as: transmission power, SNR, data rate, interference, etc. Therefore, in order to emulate a network of mobile devices we often spread nodes randomly in a terrain through simulation.

Specifically, when analyzing a cellular-based network, we usually model the Base Station (BS) coverage area by assuming simple cell shapes. There are several geometries that could be used; however it is customary to select ideal cells such as a circle or a hexagon. In fact, the hexagon is more preferred because it does not overlap with adjacent cells, nor does it skip partial surface areas [1]. Fig. 1 visually interprets the shapes.

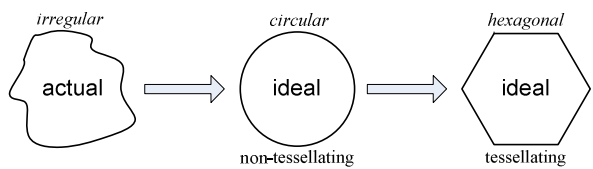

Figure 1. Different cell shapes.

Over the past two decades, since the booming of mobile telephony, many researchers looked at the effect of users' distribution on a cellular structure. In general, investigators studied the system performance through network capacity, power consumption and interference.

In particular, some contributions, such as [2] and [3], assumed that the network is circular and users were spread based on this shape. The advantage in this assumption makes the position random generation straightforward, given that the polar radial and angular Probability Density Functions (PDFs) are uncorrelated for a joint uniform distribution. On the other hand, an important drawback of this postulation is related to its non-tessellating geometry.

Other papers, such as [4–6], seem to have spread users everywhere in a hexagon, including edges. However, the distribution model was not explained thoroughly. Hence, there is no way to known for sure if the scattering was based on a truly statistical model or simply a heuristic rule that was included in the simulation code to take care of the hexagon border issue.

Further, in addition to looking at different cell contours, multiple researchers were also interested by the effect that clustering and user distributions may have on a network [4–9]. In short, they explored various nodes dropping near the BS, the edges and in an annulus within a cell. They also varied the mobile densities from uniform, to Gaussian, and sometimes to author-proposed PDFs.

In this treatment, since hexagon cells are prominent among specialists, and to our astonishment no detailed model seem to exist for random dispersion, we will start from first principle to derive one. Then, we will use a similar procedure to find a way to distribute users for sectors with rhombus and equilateral triangle shapes. Next, we will show maneuvering of clusters such as rotation and translation for convenience in dealing with large networks. While using the above, we will then demonstrate the implementation of a modular programming model that can be applied to simulate a generic complex cellular system with non-homogenous parameters such as capacity, density and sectoring. Even though, the ultimate goal in this type of research is to study the effect that terminal distributions may have on a system, it is vital to point out that in this work, we will only focus on how node positions are randomly generated for a hexagon constructed network.

## II. CELLULAR-BASED SPREADING

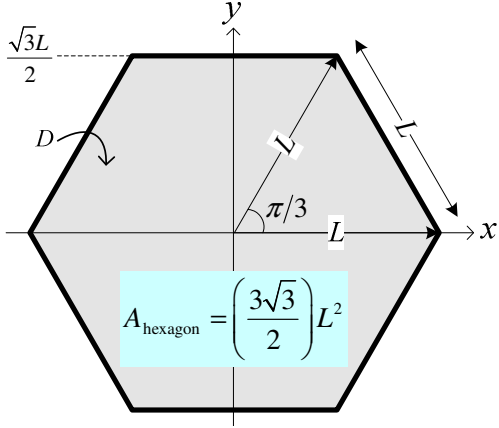

Figure 2. Hexagon cell.

Consider the hexagon shape of Fig. 2. Within this coverage area, for the sake of simplicity and perhaps from an intuitive perspective, we may very well assume that mobiles are equally spread. Because of this hypothesis, the joint PDF becomes:

$$f_{XY}(x,y) = \frac{1}{A_{hexagon}} = \frac{2}{3\sqrt{3}L^2} \quad (x,y) \in D. \quad (1)$$

In fact, if it was not for a generic structure and a priori statistical knowledge of users' trends and terrain limitations were available, then the information may have been used to ensure a more complete model. Nonetheless, using (1), we can obtain the marginal distribution for the *X*-component in (2).

$$f_X(x) = \begin{cases} \dfrac{2}{3L} & |x| \leq \dfrac{L}{2} \\ \dfrac{4}{3L^2}(L-|x|) & \dfrac{L}{2} \leq |x| < L \end{cases} \quad (2)$$

From (2), we then determine the Cumulative Distribution Function (CDF) as plotted in Fig. 3.

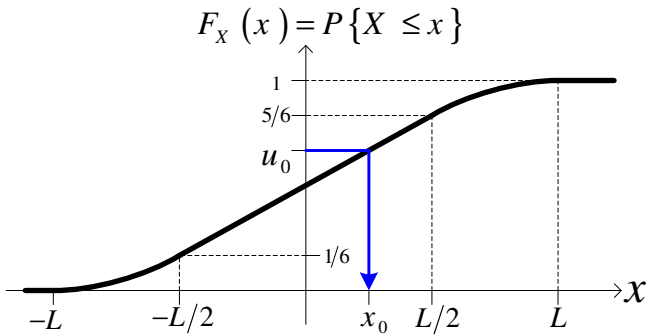

Figure 3. CDF of "*x*" for a hexagon.

Following this further, in regard to random generation, high-level computer languages such as MATLAB®, among others, have the capacity to produce a fairly long pseudorandom sequence of length $2^{1492}$ from the standard uniform distribution [10]. For any other PDF, provided the corresponding inverse CDF is available in close form, then the *Inverse Transform* method may be used [11]. Hence, we get:

$$x = F_X^{-1}(u) = \begin{cases} L\left\{\sqrt{\dfrac{3u}{2}} - 1\right\} & 0 < u \leq \dfrac{1}{6} \\ \dfrac{3L}{4}(2u-1) & \dfrac{1}{6} \leq u \leq \dfrac{5}{6} \\ L\left\{1 - \sqrt{\dfrac{3(1-u)}{2}}\right\} & \dfrac{5}{6} \leq u < 1 \end{cases} \quad (3)$$

To verify the accuracy and effectiveness of this generator, we experimented with 5,000 random points as shown in Fig. 4.

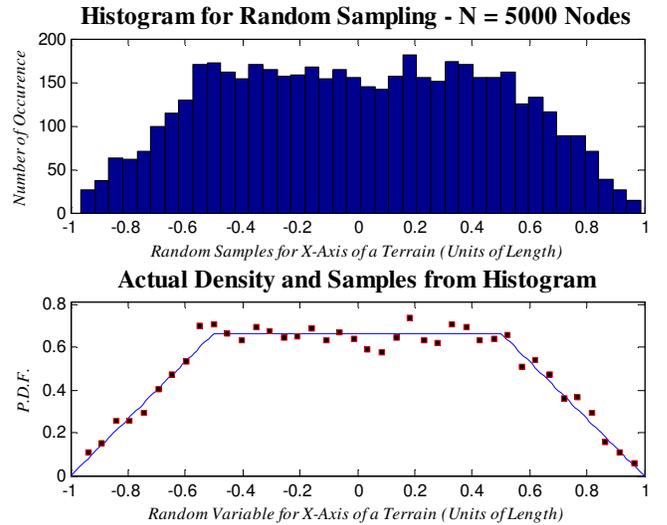

Figure 4. Random sampling from the hexagon PDF.

Further, given the obvious correlation between random variables of *X* and *Y*, and while assuming first the selection of the *X*-value, then the conditional density for *Y* becomes:

$$f_{Y|X=x_0}(y) = \begin{cases} U\left(-\sqrt{3}(x_0+L)\,;\,\sqrt{3}(x_0+L)\right) & -L < x_0 \leq \dfrac{-L}{2} \\ U\left(\dfrac{-\sqrt{3}L}{2}\,;\,\dfrac{\sqrt{3}L}{2}\right) & \dfrac{-L}{2} \leq x_0 \leq \dfrac{L}{2} \\ U\left(-\sqrt{3}(L-x_0)\,;\,\sqrt{3}(L-x_0)\right) & \dfrac{L}{2} \leq x_0 < L \end{cases}$$

(4)

Where $U(a,b) = 1/(b-a)$ is the uniform distribution over $x \in (a,b)$. As a consequence of (3) and (4), proper stochastic node scattering becomes evident as manifested by Fig. 5.

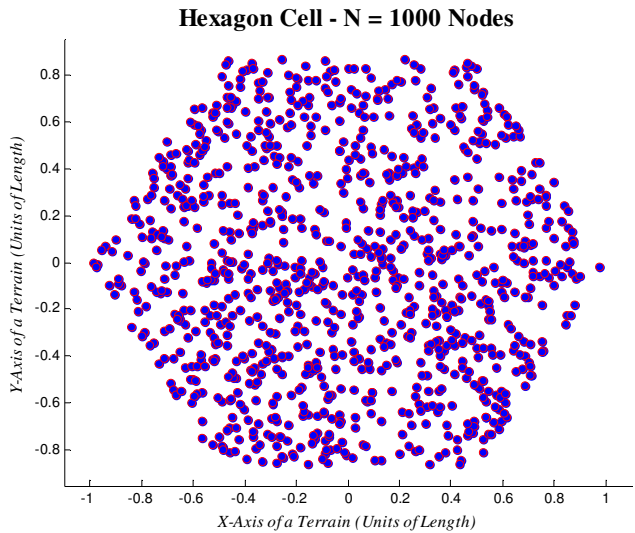

Figure 5.  Random nodes within a hexagon.

$$x = F_X^{-1}(u) = \begin{cases} \dfrac{L}{2}(2\sqrt{u}-1) & 0 < u \leq \dfrac{1}{4} \\ L\left(u - \dfrac{1}{4}\right) & \dfrac{1}{4} \leq u \leq \dfrac{3}{4} \\ L(1-\sqrt{1-u}) & \dfrac{3}{4} \leq u < 1 \end{cases} \quad (5)$$

$$f_{Y|X=x_0}(y) = \begin{cases} U\left(-\sqrt{3}x_0 \, ; \, \dfrac{\sqrt{3}L}{2}\right) & \dfrac{-L}{2} < x_0 \leq 0 \\ U\left(0 \, ; \, \dfrac{\sqrt{3}L}{2}\right) & 0 \leq x_0 \leq \dfrac{L}{2} \\ U\left(0 \, ; \, \sqrt{3}(L-x_0)\right) & \dfrac{L}{2} \leq x_0 < L \end{cases} \quad (6)$$

## III. SECTORED CELLS

Mobile systems often rely on cellular sectoring in order to ensure less interference and hence increase the number of users [1]. In general, a BS may either be omni-directional as in the previous section, with three regions, or six antennas. For example, considering a typical 120° sector, we get Fig. 6.

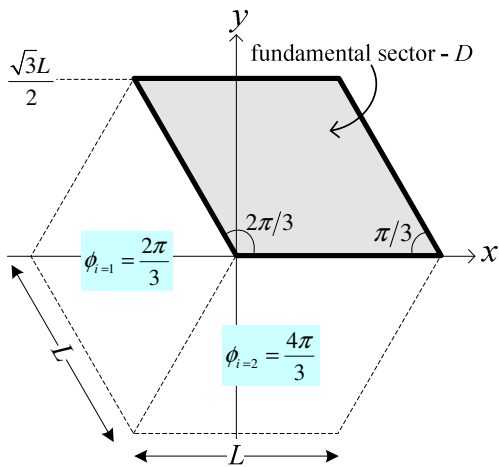

Figure 6.  Rhombus cell.

Following a similar fashion as was done for the hexagon, we obtain the inverse CDF of $X$, and the conditional density of $Y$ in (5) and (6) respectively.

Likewise for the 60° sectoring of Fig. 7, we analytically obtain (7) and (8).

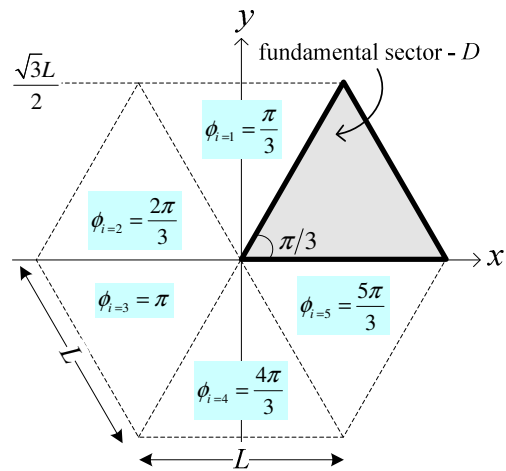

Figure 7.  Equilateral triangle cell.

$$x = F_X^{-1}(u) = \begin{cases} \sqrt{\dfrac{u}{2}}L & 0 < u \leq \dfrac{1}{2} \\ L\left(1 - \sqrt{\dfrac{(1-u)}{2}}\right) & \dfrac{1}{2} \leq u < 1 \end{cases} \quad (7)$$

$$f_{Y|X=x_0}(y) = \begin{cases} U\left(0 \, ; \, \sqrt{3}x_0\right) & 0 < x_0 \leq \dfrac{L}{2} \\ U\left(0 \, ; \, \sqrt{3}(L-x_0)\right) & \dfrac{L}{2} \leq x_0 < L \end{cases} \quad (8)$$

## IV. CLUSTER MANEUVERING

In Section III above, the mathematical expressions of (5), (6), (7), and (8) for the rhombus and the triangle are only valid within the highlighted regions shown in Fig. 6 and 7. To simulate values in the other sectors, rotation will be required.

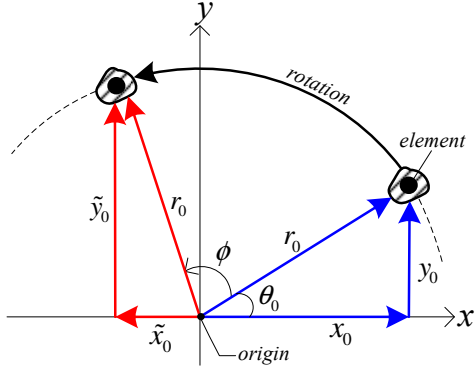

Figure 8. Element rotation.

Consider Fig. 8, where an element is rotated counterclockwise about the origin. First, if the original coordinate is expressed in polar notation and rotation is applied, then the new coordinate becomes:

$$\tilde{x}_0 = r_0 \cos(\theta_0 + \phi) = x_0 \cos(\phi) - y_0 \sin(\phi)$$
$$\tilde{y}_0 = r_0 \sin(\theta_0 + \phi) = x_0 \sin(\phi) + y_0 \cos(\phi)$$
(9)

Or if expressed differently, we find the rotation matrix:

$$\begin{bmatrix} \tilde{x}_0 \\ \tilde{y}_0 \end{bmatrix} = \begin{bmatrix} \cos(\phi) & -\sin(\phi) \\ \sin(\phi) & \cos(\phi) \end{bmatrix} \begin{bmatrix} x_0 \\ y_0 \end{bmatrix}.$$
(10)

Therefore, applying (10) with rotation angles shown in Fig. 6 and 7 will generate stochastic dots in the proper sector. Fig. 9 illustrates simulation with 1,000 nodes in each region.

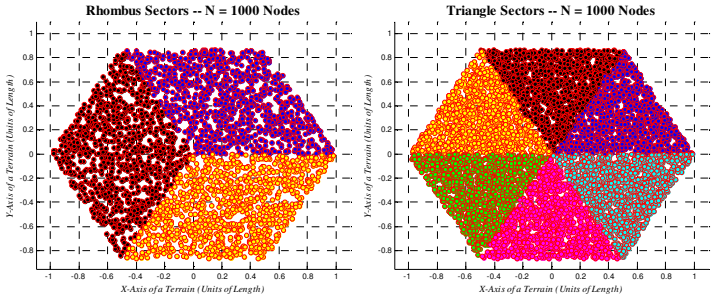

Figure 9. Shows simulation for both 120° and 60° sectoring.

Apart from rotation, large networks will have multiple cells; hence translation also becomes imperative. To help us in deriving the shifting formulas for the hexagon, we may consider the grid and the complex 19-cell network of Fig. 10.

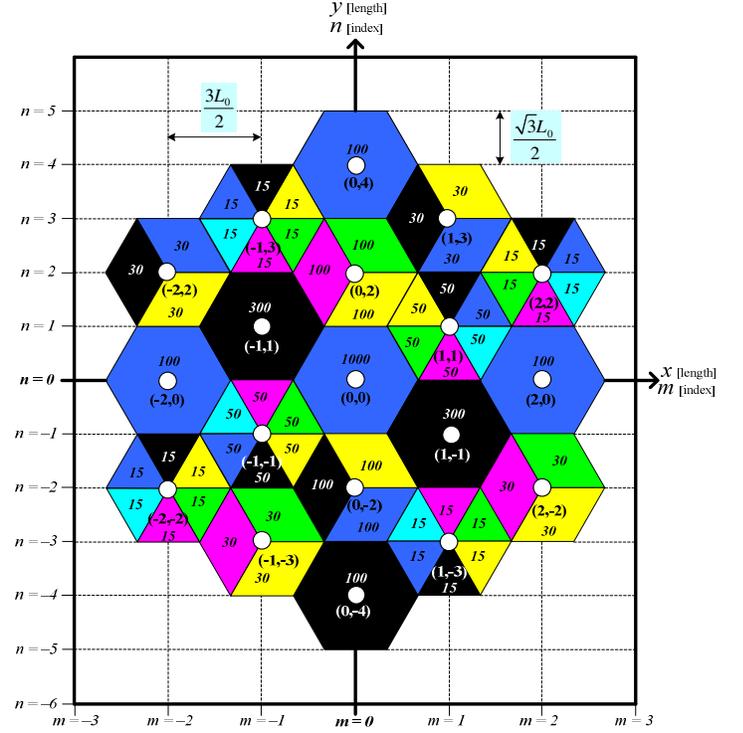

Figure 10. A complex network.

After observing the center point of the cells and applying indexing, we then find that cluster translation is given by (11) and (12), where, *mod* is the modulo operator.

$$h_m = \left(\frac{3L_0}{2}\right)m \qquad \begin{array}{l} m = 0, \pm 1, \pm 2, \cdots \\ m \in \mathbb{Z} \end{array}$$
(11)

$$k_n = \left(\frac{\sqrt{3}L_0}{2}\right)n \qquad \begin{array}{l} n = \{m \pmod{2} + 2i\} \\ i = 0, \pm 1, \pm 2, \cdots \\ (n,i) \in \mathbb{Z}^2 \end{array}$$
(12)

## V. A GENERIC SIMULATION MODEL

Now that clear and concise stochastic expressions were derived for a non-sectored and sectored hexagon, and ways to rotate and translate nodes were explained, at this point, we may tie all these concepts together in the flowchart of Fig. 11. Essentially, the block diagram shows the details that enable random spreading of nodes that can be implemented in a modular way through simulation using any computer language of choice having a standard PN sequence with *U(0,1)*.

Also, notice in the figure, that in order to cluster terminals, one would only need to input five fundamental values that describe the characteristics of a cell. In terms, to define where the cell is located, index "*m*" and "*n*" are needed. And, to determine the size of the cell, "*L*" is required. Further, to known how many sectors the cell has, "*1, 3,* or *6*" is expected.

Moreover, to identify which sector is to be considered for clustering, the proper sector *Identification (ID)* is wanted. Last, in order to establish the amount of nodes that needs to be generated, then "*N*" is also mandatory. As a result the model becomes very generic, in the sense that it can support all sort of networks easily with simple, obvious, and intuitive inputs.

quasi-stationary mobile terminals within a cellular framework. In other words, the objective was to explain methodically how random simulation is possible for a complex, dynamic, and non-homogenous network. Also, it is worth adding that despite the focus on cellular technology, this model may very well be used for relay, multihop, ad-hoc, and sensor networks to better study factors that affect the quality of service (QoS) such as system capacity, resource consumption, and interference.

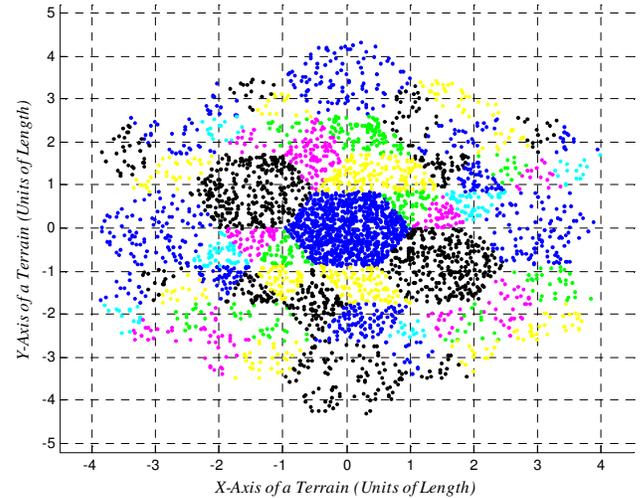

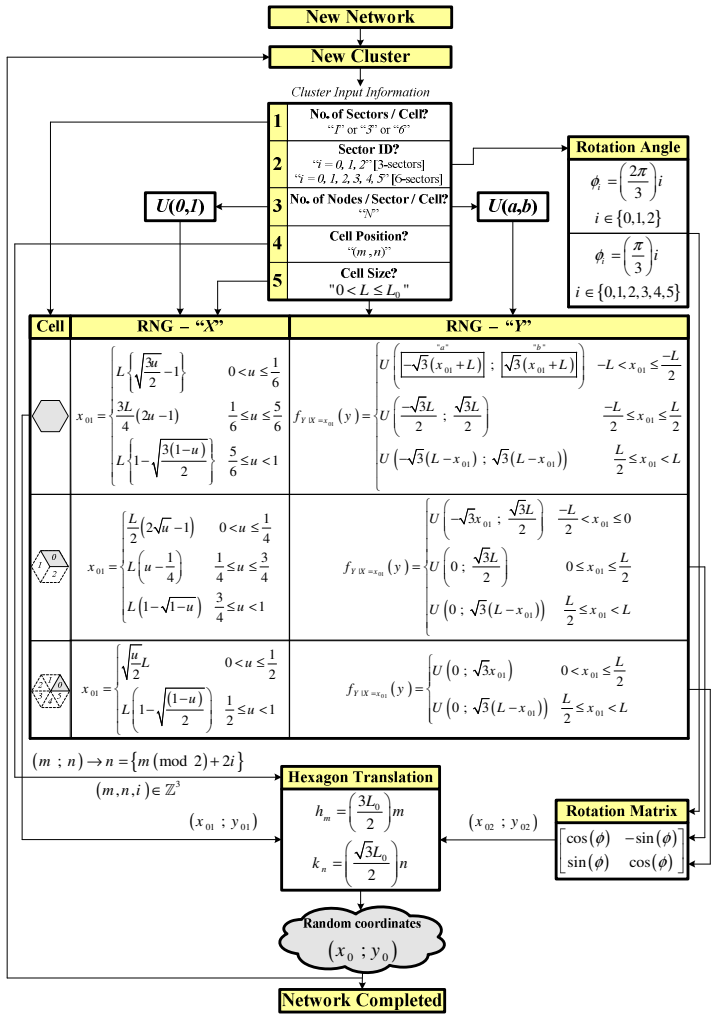

Figure 11. A generic model for cellular-based random dropping.

Figure 12. Simulation example of a complex network.

Say we simulate the network of Fig. 10, where users' density for each division is correspondingly shown in the illustration. Overall, as in Fig. 12, we have a fairly diverse network with 3920 wireless nodes. So for instance, if the interest is to look at the effect of interference, toward an origin excited terminal, then we may perhaps investigate the impact from systems in both inner and other clusters even if sectoring is used. In fact, we may conjecture that as we move away from the victim's center point, the density of systems that may negatively impact it diminishes. Hence, as reflected by Fig. 10 and 12, fewer nodes are generated deliberately in outer clusters.

## VI. CONCLUSION

It should be stressed, that the purpose of this paper was to find a coherent way to produce random points representing